\documentclass[twocolumn]{aastex62}%,linenumbers
\usepackage{amsmath,amsthm,amssymb,amsfonts}
\usepackage[utf8]{inputenc}
\usepackage[T1]{fontenc}
\usepackage{float,xspace,makecell}
\usepackage[figuresright]{rotating}
\usepackage{placeins}
\usepackage{graphicx}
\usepackage{fancyhdr}

\newcommand\mosfit{{\tt MOSFiT}\xspace}
\newcommand\mist{{\tt MIST}\xspace}

\graphicspath{{./}{figures/}}

\shortauthors{Mockler et al. 2021}
\begin{document}
%\pagestyle{fancy}

%\fancyhf{}
%\markcenter{ UC Presidents Postdoctoral Fellowship -- Education Background \& Personal Experience -- Brenna Mockler}
%\fancyhead[LE,LO]{\textsl{Mockler, Brenna - Writing Sample}}

\title{Evidence for the preferential disruption of moderately massive stars by supermassive black holes} 

\author[0000-0001-6350-8168]{Brenna Mockler}
\email{bmockler@ucsc.edu}
\affiliation{Department of Astronomy and Astrophysics, University of California, Santa Cruz, CA 95064, USA}

\author{Angela A. Twum}
\affiliation{Department of Astronomy and Astrophysics, University of California, Santa Cruz, CA 95064, USA}

\author[0000-0002-4449-9152]{Katie Auchettl}
\affiliation{School of Physics, The University of Melbourne, VIC 3010, Australia}
\affiliation{ARC Centre of Excellence for All Sky Astrophysics in 3 Dimensions (ASTRO 3D)}
\affiliation{Department of Astronomy and Astrophysics, University of California, Santa Cruz, CA 95064, USA}
\affiliation{OzGrav, University of Melbourne, Parkville, Victoria 3010, Australia}

\author[0000-0002-3696-8035]{Sierra Dodd}
\affiliation{Department of Astronomy and Astrophysics, University of California, Santa Cruz, CA 95064, USA}

\author[0000-0002-4235-7337]{K.D. French}  
\affiliation{Department of Astronomy, University of Illinois, 1002 W. Green St., Urbana, IL 61801, USA}  

\author[0000-0001-8825-4790]{Jamie A.P. Law-Smith}
\affiliation{Center for Astrophysics $|$ Harvard \& Smithsonian, 60 Garden St., Cambridge, MA 02138, USA}

\author[0000-0003-2558-3102]{Enrico Ramirez-Ruiz}
\affiliation{Department of Astronomy and Astrophysics, University of California, Santa Cruz, CA 95064, USA}

\correspondingauthor{Brenna Mockler}

\begin{abstract}
Tidal disruption events (TDEs) provide a unique opportunity to probe the stellar populations around supermassive black holes (SMBHs). By combining light curve modeling with spectral line information  and knowledge  about the stellar populations in the host galaxies, we are able to constrain the properties of the disrupted star for three TDEs. The TDEs in our sample have UV spectra, and measurements of the UV \ion{N}{3} to \ion{C}{3} line ratios enabled estimates of the nitrogen-to-carbon abundance ratios for these events. We show that the measured nitrogen line widths are consistent with originating from the disrupted stellar material dispersed by the central SMBH. We find that these nitrogen-to-carbon abundance ratios necessitate the disruption of moderately massive stars ($\gtrsim 1 - 2 M_\odot$). We determine that these moderately massive disruptions are over-represented by a factor of $\gtrsim 10^2$ when compared to the overall stellar population of the post-starburst galaxy hosts. This implies that SMBHs are preferentially disrupting higher mass stars, possibly due to ongoing top-heavy star formation in nuclear star clusters or to dynamical mechanisms that preferentially transport higher mass stars to their tidal radii.

\end{abstract}

\section{Introduction} \label{sec:intro}
The gravitational influence of a super-massive black hole dominates the kinematics of stars within a nuclear cluster. Each star within this dense region traces out an intricate trajectory under the combined influence of the SMBH and the other stars. Stellar dynamics is capable of describing the resulting stellar motions as point masses moving under the influence of gravity until these encounters move a star onto a nearly radial orbit. If a star wanders close to the tidal radius it is ripped apart by the SMBH's gravity \citep{Rees:1988a}. When that happens, the internal structure of the star has to be taken into consideration and we must switch to a hydrodynamic description to follow the evolution of the disrupted stellar debris. A fraction of the debris eventually falls back, circularizes, and accretes onto the SMBH. This accretion powers a flare which is a clear sign of the presence of an otherwise quiescent SMBH. As we show here, it can also provide a compelling fingerprint of the properties of the disrupted star. 

It has only been in the last decade -- with the advent of numerous wide field transient surveys -- that we have started to collect photometric and spectroscopic data on a myriad of tidal disruption events \citep[TDEs; e.g., ][and references therein]{van-Velzen:2020}. The discovery of TDE flares has generated widespread excitement, as we can use them to study the masses of SMBHs in quiescent galaxies \citep{Mockler:2019, Wen:2020, Ryu:2020a}, the populations and stellar dynamics in galactic nuclei \citep{Kochanek:2015a, Kochanek:2016a, Yang:2017}, and the physics of black hole accretion \citep{Dai:2018, Bonnerot:2020a, Andalman:2020}. The observed rates of TDEs hold important discriminatory power over both the dynamical mechanisms operating in galactic nuclei and the nature of their underlying stellar populations. Both the underlying stellar populations and the dynamical mechanisms that feed stars into disruptive orbits remain uncertain, particularly after the surprising observation that TDEs preferentially occur in an unusual sub-type of galaxies known as E+A galaxies \citep{Arcavi:2014a, French:2016a, Law-Smith:2017b}.

Models of tidal disruption light curves can be used to estimate the parameters of disruption, such as the mass of the black hole and star, and the efficiency of the conversion from accreted mass to observed luminosity \citep{Mockler:2019, van-Velzen:2019b, Ryu:2020a, Zhou:2021}. However, when modeling the light curve, there is a large degeneracy between the efficiency of conversion from mass to energy and the mass of the disrupted star \citep{Mockler:2021a}. 

Helping constrain the nature of disrupted stars can break the degeneracy between the mass and efficiency as well as provide invaluable insights into the stellar populations and the dynamical processes in galactic nuclei. Recent measurements of metal lines in the UV spectra of a subset of TDEs has provided us with a unique opportunity to uncover the metal content of the reprocessing material. Assuming that the emitting gas originates from the stellar debris, the mass and evolutionary state of the disrupted star can be constrained independently. Measurements of nitrogen-to-carbon UV line ratios are particularly useful for estimating the  corresponding abundances in the emitting gas. As the CNO cycle creates a surplus of nitrogen-rich and carbon-deficient material \citep{Kochanek:2015a,Gallegos-Garcia:2018,Law-Smith:2019}, TDEs from evolved stars could naturally explain enhanced N/C abundance ratios as derived from $\rm N^{2+}/C^{2+}$ line ratios. This argument is particularly robust when considering the UV  \ion{N}{3} and \ion{C}{3} lines, as these elemental transitions have similar critical densities and excitation energies, and involve ions with similar ionization energies \citep{Yang:2017}. 

In this paper we focus on ASASSN-14li \citep{Holoien:2016a, Cenko:2016a,Brown:2016b}, iPTF15af \citep{Blagorodnova:2019,Onori:2019}, and iPTF16fnl \citep{Blagorodnova:2017a, Brown:2017a}, three TDEs whose spectra all have relatively high nitrogen-to-carbon abundance ratios as derived from UV \ion{N}{3}/\ion{C}{3} line ratios \citep{Yang:2017}. For completeness we also include an analysis of ASASSN-18pg (AT2018dyb), which has particularly prominent nitrogen lines in its optical spectra, but no UV spectra, and therefore the corresponding  N/C  abundance ratios are more difficult to constrain. 

This paper is structured as follows.
In Section~\ref{sec:lightcurves} we estimate the location of the nitrogen enriched gas as derived by the equivalent line widths and compare it to the expected location of debris from the disrupted star in order to show whether or not the emitting gas could have originated from the undigested stellar matter as opposed to material present before the flare. In Section~\ref{sec:composition} we show that the super-solar nitrogen abundances single out the disruption of moderately  massive ($\gtrsim 1-2 M_{\odot}$) stars. To understand this over-representation of moderately massive (or super-solar) stars, in Section~\ref{sec:hostgalaxies}, we study the host galaxies of these TDEs. The host galaxies of all three TDEs are E+A or quiescent Balmer-strong galaxies, suggesting a recent episode of star formation. Finally, in Section~\ref{sec:disc} we present our conclusions and discuss several theories that might explain the over-abundance of high mass disruptions.  

\begin{figure}[ht!]
\begin{center}
\includegraphics[scale = 0.48]{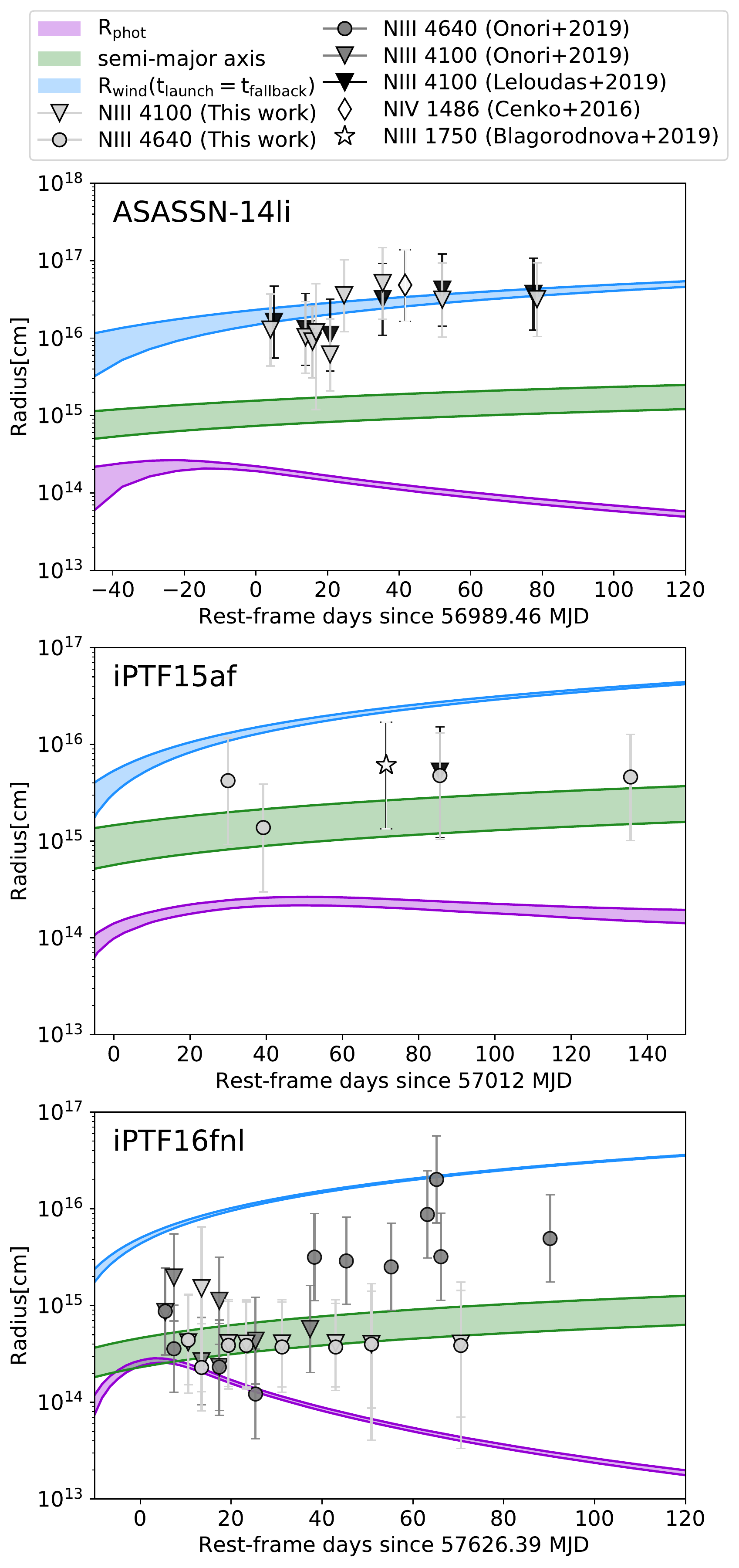}
\end{center}
\vspace{-0.55cm}
\caption{ The various size scales of gas around the black hole originating from the disrupted star. The blackbody photosphere is plotted in purple, the semi-major axis of debris returning at a given time is plotted in green, and the wind radius (assuming a constant, 0.1c velocity wind launched at first fallback) is plotted in blue. We note that the semi-major axis plotted is a minimum -- at any given time, debris that is still orbiting will have larger semi-major axes than debris that has just returned to the black hole. The nitrogen lines appear to emanate from between the photosphere and the wind radius, consistent with originating from the disrupted star. The lines for ASASSN-14li are furthest from the SMBH, but consistent with a wind scenario (see text).
\label{fig:blr}}
%\vspace{-0.5cm}
\end{figure}

\section{LESSONS LEARNED FROM LIGHT CURVES AND SPECTRA OF TDE\lowercase{S}}\label{sec:lightcurves}
The broad line region (BLR) in many AGN is thought to be produced by gas at distances from light hours to light years away from the SMBH \citep{Blandford:1982,Vestergaard:2002, McLure:2002}. This wide range of scales is in contrast to the debris disk assembly following a TDE. Rather than gas spiraling in from sub-parsec scales, the debris disk forms from the inside out and has an initial characteristic scale of tens of AU \citep{Guillochon:2014a}. Despite these differences, similar physical mechanisms to those operating in steady-state AGN may also be at work in TDEs.  While it is still debated whether the material responsible for the BLR is in the form of a disk wind or clouds, the irradiated gas is thought to be bound to the SMBH with information on the structure and kinematics of this region commonly derived from the broad emission line profiles \citep[e.g.,][]{Pancoast:2014a,Pancoast:2014b, Pancoast:2018}. 

Through comparison with the processes responsible for producing the BLR of steadily-accreting AGN together with the commonly held belief that the line width FWHM is related to the rotational motion of the gas, in this Section we infer the location of the nitrogen emitting gas and investigate whether the TDE debris has sufficient time to reach this distance by the time the first spectrum was observed in these events. 

Light curve fitting provides rough constraints on the evolution of stellar debris surrounding the SMBH. By estimating the blackbody photosphere, the orbits of in-falling debris and the possible launch times for winds with the \mosfit transient fitting code\footnote{the publicly \href{https://mosfit.readthedocs.io/en/latest/}{available} Modular Open Source Fitter for Transients (\mosfit) uses Markov-Chain Monte Carlo processes to fit analytical and semi-analytical models to multi-band transient light curves, extracting the most likely parameter distributions \citep{Guillochon:2018}.}, we are able to estimate the size scales of gas structures expected to exist around the SMBH after the TDE occurs \citep{Mockler:2019}. We can then compare these size scales to the virial radii estimated using the FWHM of the nitrogen lines in the TDE spectra. We measured the FWHM for the \ion{N}{3} $\lambda$4100 and \ion{N}{3} $\lambda$4640 lines for all events included in this work (see Appendix~\ref{app:fwhm}), and also included the values measured previously in the literature in Figure~\ref{fig:blr}. As there is often significant uncertainty in measuring the line widths, including multiple measurements gives a better picture of the systematic uncertainty. 

Hydrodynamic simulations of stellar disruptions show the bound debris spreading out in elliptical orbits with size scales dependent on the relative binding energy of the debris to the black hole \citep{Guillochon:2014a}. If the nitrogen-enriched gas does originate from the disrupted star, we expect it to lie somewhere between the blackbody photosphere radius and the outer edge of an accretion-driven or collisionally-induced wind launched by the return of the most bound material. Most TDEs have peak luminosities within a factor of a few of their Eddington limit \citep{Mockler:2021a}, and these high luminosities are expected to power strong winds, which efficiently carry mass far from the SMBH \citep{Dai:2018}.  Additionally, simulations of accretion disk formation show that the super-Eddington mass inflow rates suggest significant mass ejection \citep{Ramirez-Ruiz:2009a}, and the circularization process can also produce strong, shock-driven winds reaching velocities between $\approx 0.01-0.1$c \citep{Lu:2020a}. 

In Figure~\ref{fig:blr} we plot the virial radii of the nitrogen gas alongside the blackbody photosphere evolution, the semi-major axis of the orbiting bound debris (which increases with time), and the outer edge of a wind launched at a velocity of 0.1c at the time of first fallback (when debris first returns to the black hole after the star is disrupted). We find that the locations of the nitrogen lines are consistent with debris originating from the disrupted star and its inferred location lies between the photosphere radius and the edge of a fast wind. The nitrogen lines for ASASSN-14li are the farthest from the SMBH, yet consistent with the a wind scenario. This is consistent with other data and modeling of ASASSN-14li. \citet{Alexander:2016a} and \citet{Kara:2018} measured a wind moving at $\gtrsim$0.1c, and \citet{van-Velzen:2016a} suggest the decay rate of the radio luminosity matches that of a decelerating, mildly relativistic wind.

We note that it is possible the line broadening is due to transport effects instead of kinematics. For example, \citet{Roth:2018} found that line transport through gas with high optical depths to electron scattering could also broaden lines in TDEs and reproduce observed FWHM velocities. If the measured FWHM values are due to electron scattering or other transport effects and not kinematics, the lines would have to originate at smaller radii (near the edge of the photosphere). This would reduce the similarity with AGN broad line regions, however, the gas could still originate from the stellar debris.

\section{Lessons learned from composition}\label{sec:composition}

The events included in this paper are part of a sub-class of ``nitrogen-rich'' TDEs that has been assembled in the last few years \citep{Brown:2017a, Cenko:2016a, Blagorodnova:2019, Holoien:2020, Hung:2017a, Hung:2019, Hung:2020, Hung:2021, van-Velzen:2021}. These events have broad optical \ion{N}{3} lines in addition to the more commonly observed He and H emission lines. For at least some events in this sub-class, the \ion{N}{3} lines have been theorized to be due to the super-solar nitrogen abundances present in the material of disrupted stars, which is then dispersed and subsequently irradiated by the central SMBH. The reader is referred to Section~\ref{sec:disc} for a discussion on the demographics of TDEs and the fraction of events with clear nitrogen signatures.

\begin{figure}[h!]
\includegraphics[width=\columnwidth]{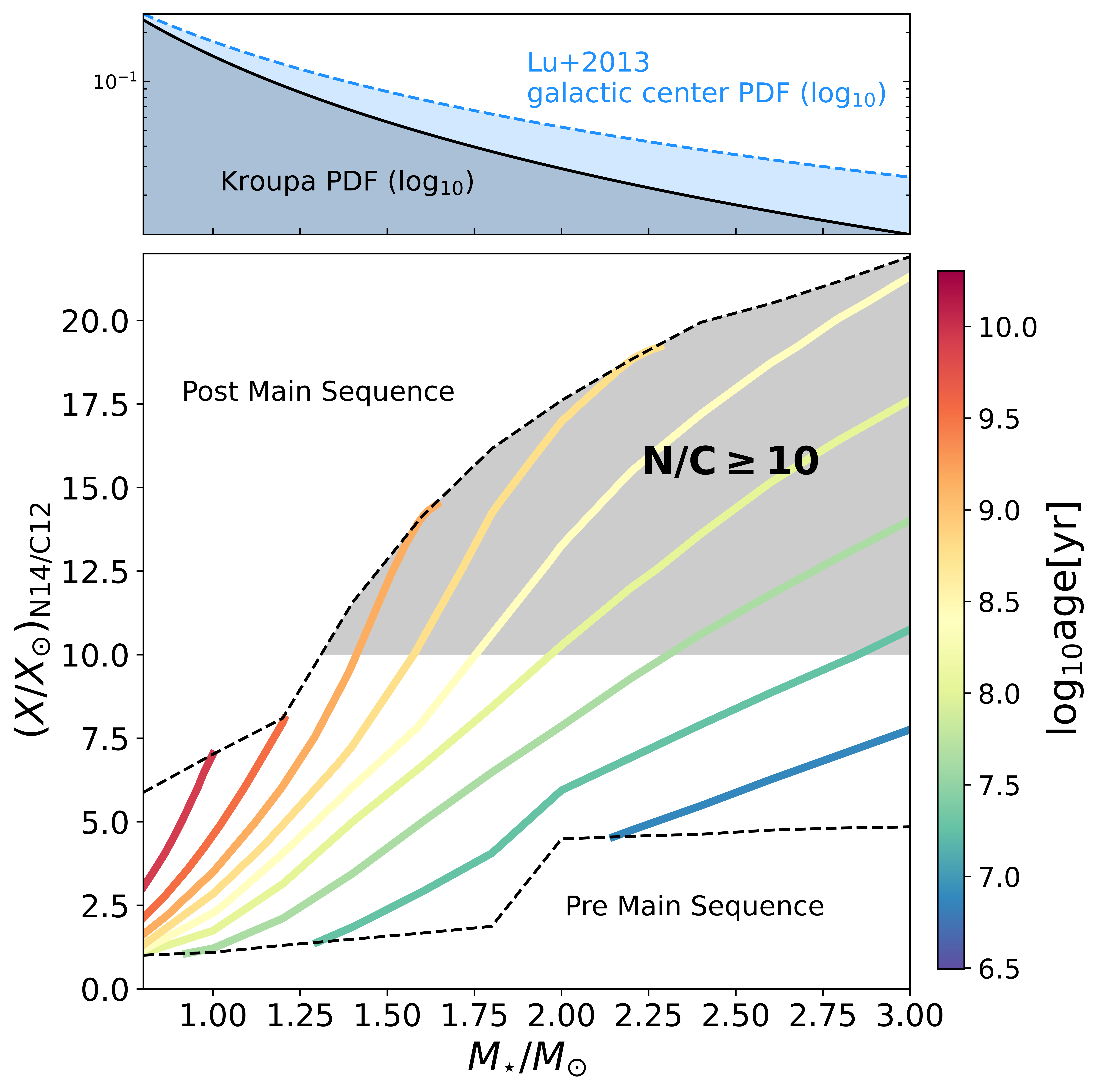}
\vspace{-0.3cm}
\caption{Mass versus N/C composition for the tidal debris of main sequence stars. Lines of constant age are plotted in rainbow colors. The shaded gray region denotes stars with N/C $\geq 10$. The smallest stars that reach this N/C abundance are 1.3 $M_\odot$. We focus on stars below $3 M_\odot$ as they are much more prevalent, however we include higher mass stars in our calculations. We use the fallback models calculated by \citet{Gallegos-Garcia:2018}, and plot the value of the composition at $\rm t_{fallback} = 0.1 \times t_{fallback, \; peak}$.}
\label{fig:mass-comp}
\end{figure}

The connection between the line strengths and the relative abundances is dependent on the gas conditions where the lines are produced, which are difficult to measure and likely vary within the line-emitting regions as the density and temperature changes with distance to the SMBH. This is the case, for example, when trying to interpret the helium-to-hydrogen line ratios as an indication of the disruption of helium enhanced stars \citep{Roth:2016a}. 

Measurements of nitrogen-to-carbon UV line ratios are, on the other hand, particularly robust for estimating the  corresponding abundances in the emitting gas. This is because the \ion{N}{3} ($\lambda 1750$) and \ion{C}{3} ($\lambda 1908$) line transitions have similar critical densities and excitation energies, and involve ions with similar ionization energies \citep{Yang:2017}. This means that the line ratio of \ion{N}{3}/\ion{C}{3} does not depend strongly on the gas conditions of individual TDEs, unlike optical line N strengths. As such, the UV line ratio \ion{N}{3}/\ion{C}{3} can be used to estimate the abundance ratio of N/C. While a number of TDEs have broad nitrogen lines measured in their optical spectra, the three events discussed in this paper (ASASSN-14li, iPTF15af, and iPTF16fnl) are the only TDEs that have nitrogen lines measured in their UV spectra that can be used to robustly estimate the N/C abundance ratios. All three events occurred relatively close by ($z \leq 0.08$), making them good candidates for HST spectroscopy. They have luminosities and black hole masses consistent with the general population of observed TDEs. We note that there are 10 TDEs with broad nitrogen lines measured in their optical spectra that do not have HST UV spectra. For comparison, we analyze one of these events (ASASSN-18pg) in Appendix~\ref{app:18pg}.

In \citet{Yang:2017}, the authors used the UV \ion{N}{3} and \ion{C}{3} line ratios measured for ASASSN-14li, iPTF15af, and iPTF16fnl to estimate N/C abundances. They explored the dependence of the abundance ratio calculation on parameters such as the input spectral energy distributions (SEDs), the gas density, and the ionization parameter ($U$).

\citet{Yang:2017} found a strict lower limit of N/C$\geq$10 for ASASSN-14li. This event has the lowest N/C line ratio of any of the three transients, and because of the insensitivity of the line ratio to both $U$ and the shape of the SED, it is argued by \citet{Yang:2017} to provide a reasonable lower limit for iPTF15af and iPTF16fnl as well.\footnote{ASASSN-14li was brighter than iPTF16fnl, and comparable in luminosity to iPTF15af \citep{Holoien:2016a,Jiang-N:2016,Blagorodnova:2017a,Blagorodnova:2019}.} 

The constraints derived on the N/C composition of the stellar debris associated with ASASSN-14li, iPTF15af, and iPTF16fnl can help constrain the masses of the disrupted stars. In Figure~\ref{fig:mass-comp}, we plot the stellar mass as a function of the fallback debris composition for main sequence stars at a range of stellar masses and ages. We use the composition dependent mass fallback rates calculated in \citet{Gallegos-Garcia:2018} to estimate the maximum N/C abundance ratios that could reasonably be present at the time the UV spectra were taken. The N/C abundance ratio in the debris increases with time, as material that originated closer to the core of the star with higher N/C abundance ratios is less bound to the black hole and takes longer to return. 

In our calculations, we use the composition of the debris at the time the fallback rate drops to 0.1 of its peak value, which occurs at approximately 10 peak timescales for most full disruptions. This is after the UV spectra were taken; however, we want to be mindful of possible mixing across stellar layers. Hydrodynamical simulations show that mixing can decrease the time for nitrogen-enhanced debris to return to the black hole by a factor of $\approx 2-4$ \citep{Law-Smith:2019}. All of the UV HST measurements used in this paper were taken within $\approx 2$ peak timescales after disruption for their respective transients, and so it is reasonable to use the N/C abundance derived by \citet{Gallegos-Garcia:2018} after 10 peak timescales as a strict upper limit to the abundance and, correspondingly, a strict lower limit to the mass of the disrupted star.  

Using these models of the composition of fallback debris we find, as expected, that low mass stars at all ages never reach high enough N/C abundance ratios to be compatible with the observed line ratios. We find that the minimum stellar mass with a high enough N/C ratio is  $\approx 1.3 M_\odot$ (see the gray shaded region in Figure~\ref{fig:mass-comp}), which provides us with a strict lower limit. Yet, stars with masses close to $1.3 M_\odot$ would need to be near the end of their main sequence lifetimes to have N/C$\geq$10. Stars with masses equal to or above $1.3 M_\odot$ are relatively rare -- we have plotted the Kroupa IMF and the \citet{Lu:2013} galactic center IMF in the top panel of Figure~\ref{fig:mass-comp}. In addition to being less common at birth, these stars also have shorter lifetimes than their lighter siblings. Therefore, as the nuclear stellar population ages, high mass stars become subsequently less common if they are not effectively replenished. In Section~\ref{sec:hostgalaxies} we analyze the age distributions of the stars residing in the corresponding host galaxies and estimate the disruption probability of stars with N/C$\geq$10 under the assumption that such populations might be representative of the stellar populations present in the nuclei of the galaxies. The validity of this assumption is revisited in Section~\ref{sec:disc}.

\section{Lessons learned from host galaxies}\label{sec:hostgalaxies}

\begin{figure*}
\begin{center}
\includegraphics[width=6.5in]{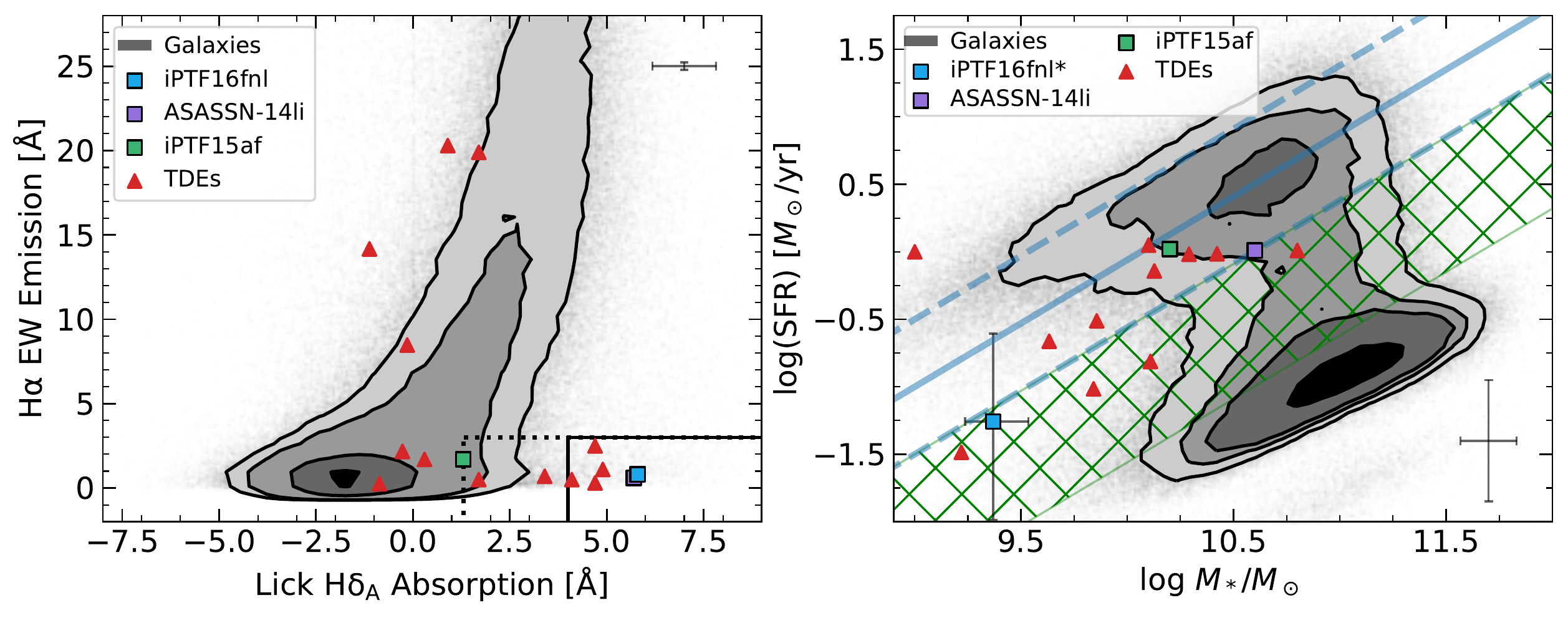}
\vspace{-0.3cm}
\caption
{The gray contours represent galaxies from the SDSS reference catalog described in \citet{Law-Smith:2017b}. TDE host galaxies are represented by red triangles, with our sample of 3 TDEs plotted separately using colored squares. {\bf Left panel:} H$\alpha$ equivalent width as a function of  Lick H$\delta$A absorption for TDES including those studied in this paper \citep{French:2016a, Law-Smith:2017b, French:2020a, Dodd:2021}. Galaxies inside the solid line box are categorized as E+A/PSB galaxies while galaxies inside the dotted line box are defined as QBS galaxies. {\bf Right panel:} the star formation rate as a function of the total stellar mass. The star forming main sequence is plotted as a blue solid line, the dashed lines represent $\pm 0.5$dex, the scatter in the star formation rate measurements used in this plot. Following \citet{Law-Smith:2017b} and \citet{Pandya:2017}, we define the green valley as the region between $1\sigma$ and $3\sigma$ below the star forming main sequence (plotted in green hatching).
}
\end{center}
\vspace{-0.2cm}
$^*${\footnotesize The SFR for iPTF16fnl has not been measured, and so to estimate a range of likely values we took the range of SFR's for galaxies in our catalog with similar values of $\rm H\alpha EW$, $\rm H\delta_A$ and stellar mass. In particular, we included galaxies within $1\sigma$ in stellar mass and $2\sigma$ in $\rm H\alpha \; EW$ and $\rm H\delta_A$ (as these galaxies are rare, and there were no galaxies within $1\sigma$).}\\
\label{fig:hostgal}
\end{figure*}
 
\subsection{Host galaxy types}

\citet{Arcavi:2014a} and \citet{French:2016a} reported on the surprising observation that TDEs preferentially occur in an unusual sub-type of galaxies known as E+A galaxies. E+A (or K+A) galaxies are so called due to Balmer absorption features in their spectra (characteristic of an A star) which appear superimposed on an (E)arly-type galaxy population  \citep[or old K star;][]{Dressler:1983a}. These galaxies are rare: they comprise only 0.2\% percent of the local population, yet host 20\% of the optically and UV-detected TDE candidates with measured host galaxy properties to date \citep{Law-Smith:2017b, French:2020a}. The Balmer absorption in E+A galaxies points to a significant starburst population with ages $10^8-10^9$ years, while low H$\alpha$ indicates a lack of ongoing star formation (Figure~\ref{fig:hostgal}). For this reason, they are also called post-starburst (PSB) galaxies. Galaxies with slightly weaker Balmer absorption, likely due to a smaller recent star formation episode, are generally classified as quiescent Balmer-strong (QBS) galaxies.  Following \citet{French:2020a}, we define galaxies with $\rm H\delta_A - \sigma(H\delta_A) >4$\r{A} as E+A/post-starburst and galaxies with $\rm H\delta_A>1.31$\r{A} as quiescent Balmer-strong. In both cases we require $\rm H\alpha \; EW < 3$\r{A}. Interestingly, the hosts of the three  TDEs with super-solar nitrogen abundances studied here are all quiescent Balmer-strong galaxies, with two of them being E+A or post-starburst. For simplicity we will refer to all of these galaxies with recent starbursts as `post-starburst', except when the distinction is important to the meaning of the text. TDE host galaxies also have relatively high central concentrations of stellar light, as measured by the steepness of their brightness profiles \citep{Law-Smith:2017b}, which is commonly argued to roughly indicate centrally enhanced stellar densities.

As such, nitrogen-rich (N/C $\geq 10$) TDEs require the disruption of moderately massive ($\geq 1.3 M_\odot$) stars, and they appear to occur in the centers of galaxies where two conditions are met: the presence of a relatively recent starburst episode and a large concentration of stellar mass in the central regions. The mechanism by which these moderately massive stars in galactic nuclei are ferried to the very close vicinity of the SMBH remains an open question. There have been several theories put forward to explain the enhanced TDE rate in post-starburst galaxies, from an over-density of stars close to the SMBH \citep{Law-Smith:2017b, Stone:2018a}, to the presence of SMBH binaries \citep{Chen:2011a,Arcavi:2014a, Li:2015a}, to star formation in eccentric disks around SMBHs \citep{Madigan:2018}. What these theories share in common, is that the disrupted stars come from within the radius of influence of the SMBH. Deep in the potential of the black hole, stars gravitationally interact with one another coherently, in contrast to two-body relaxation, resulting in rapid angular momentum evolution \citep{Rauch:1996a}. This rapid evolution allows for moderately massive stars to be disrupted before they can significantly evolve. In what follows we assume that the stellar populations of galaxy nuclei resembles that of the host in order to make inferences about the rate enhancement required to explain the inferred N/C abundance ratios.

\subsection{Stellar population content in TDE hosts}\label{sec:stellarpop}
To estimate the number of stars that can produce nitrogen-rich tidal disruptions (defined here as tidal debris with N/C$\geq 10$), we  first calculate the relative number of moderately massive main sequence stars present in the corresponding hosts. As shown in Figure~\ref{fig:mass-comp}, if we can calculate the masses and ages of the stellar population, we can determine what percentage of the stellar population could have a significant nitrogen enhancement if disrupted. As all three of the events in our sample reside in post-starburst galaxies, we investigate whether the recent starbursts can help explain the disruption of moderately massive  stars.

We make use of the star formation histories derived in \citet{French:2017} for the host galaxies of ASASSN-14li and iPTF15af, together with a Kroupa initial mass function \citep{Kroupa:2001} to get the masses and ages of the overall stellar populations in these galaxies (we also vary the IMF to better represent galactic nuclei, as described below). We do a separate analysis of the stellar population in iPTF16fnl, as this galaxy does not have as detailed of constraints on its star formation history. Following \citet{French:2017}, we use
\begin{align*}
     \Psi &\propto t e^{-t/\tau_{\rm o}}; & t < 10^{10} - t_{\rm y} \\
    &\propto t e^{-t/\tau_{\rm o}} + \beta e^{-(t + t_{\rm y} - 10^{10})/\tau_{\rm y}}; & t \geq 10^{10} - t_{\rm y},
\end{align*}
to jointly model the old stellar population's star formation history combined with the young stellar population's recent starburst. We assume a linear-exponential star formation rate for the old stellar population, beginning 10 Gyr ago ($t=0$) with a characteristic timescale $\tau_{\rm o}=1$Gyr. The young starburst is modelled as an exponential decline in star formation rate that started $t_{\rm y}$ years ago with a characteristic timescale $\tau_{\rm y}$ as determined by fits to spectroscopic observations. The normalization factor between the old and young starburst, $\beta$, ensures the total fraction of stars from the young starburst is consistent with the corresponding burst mass fraction determined from observations. The host of ASASSN-14li (iPTF15af) has $t_{\rm y} = 473^{+373}_{-67}$ Myr ($t_{\rm y} = 595^{+108}_{-191}$ Myr), $\tau_{\rm y}=25-200$Myr ($\tau_{\rm y}=25-100$Myr) and a burst mass fraction $=0.055^{+0.016}_{-0.016}$ (burst mass fraction $=0.005^{+0.002}_{-0.002}$), as derived by \citet{French:2017}.

We then use the relative likelihood of disruption based on each star's mass and radius. Higher mass stars and more evolved stars are slightly easier to disrupt, as they are less dense, which translates to a more extended tidal radii, $R_{\rm t}$. We use the rate estimate described in \citet{MacLeod:2012a} to determine the corresponding mass weighting: 
\begin{equation}\label{eqn:rate}
    \frac{dr}{dM_*} \propto 
    R_{\rm t}^{1/4} \frac{dn}{dM_*} \propto \frac{R_*^{1/4}}{M_*^{1/12}} 
    \frac{dn}{dM_*}
\end{equation}

As described above, the stellar population, $dn/dM_*$, is calculated using the star formation rates estimated in \citet{French:2017}. We use \mist \citep{Dotter:2016,Choi:2016,Paxton:2011, Paxton:2013,Paxton:2015} to calculate the mass-radius relation as a function of stellar mass and age (with the zero age main sequence metallicity set to solar metallicity using the \citet{Asplund:2009} value of Z = 0.0142). For simplicity we first proceed under the assumption that the diffusion coefficients for a given SMBH environment are mass independent and that any changes in the tidal disruption rate arises solely based on differences in the mass-dependent cross section (see Section~\ref{sec:disc} for further discussion). 

\subsection{On the preferential disruption of  $\geq 1.3 M_\odot$ stars}
After weighting the stellar populations by the relative likelihood of disruption as a function of stellar mass, we are able to integrate over the range of stellar masses and ages that can produce the observed N/C line ratios. As such, we can calculate the fraction of stars that satisfied these constraints under the assumption that the overall stellar population of the host is representative of that of the nuclei. 

\begin{figure}[h!]
\includegraphics[width=\columnwidth]{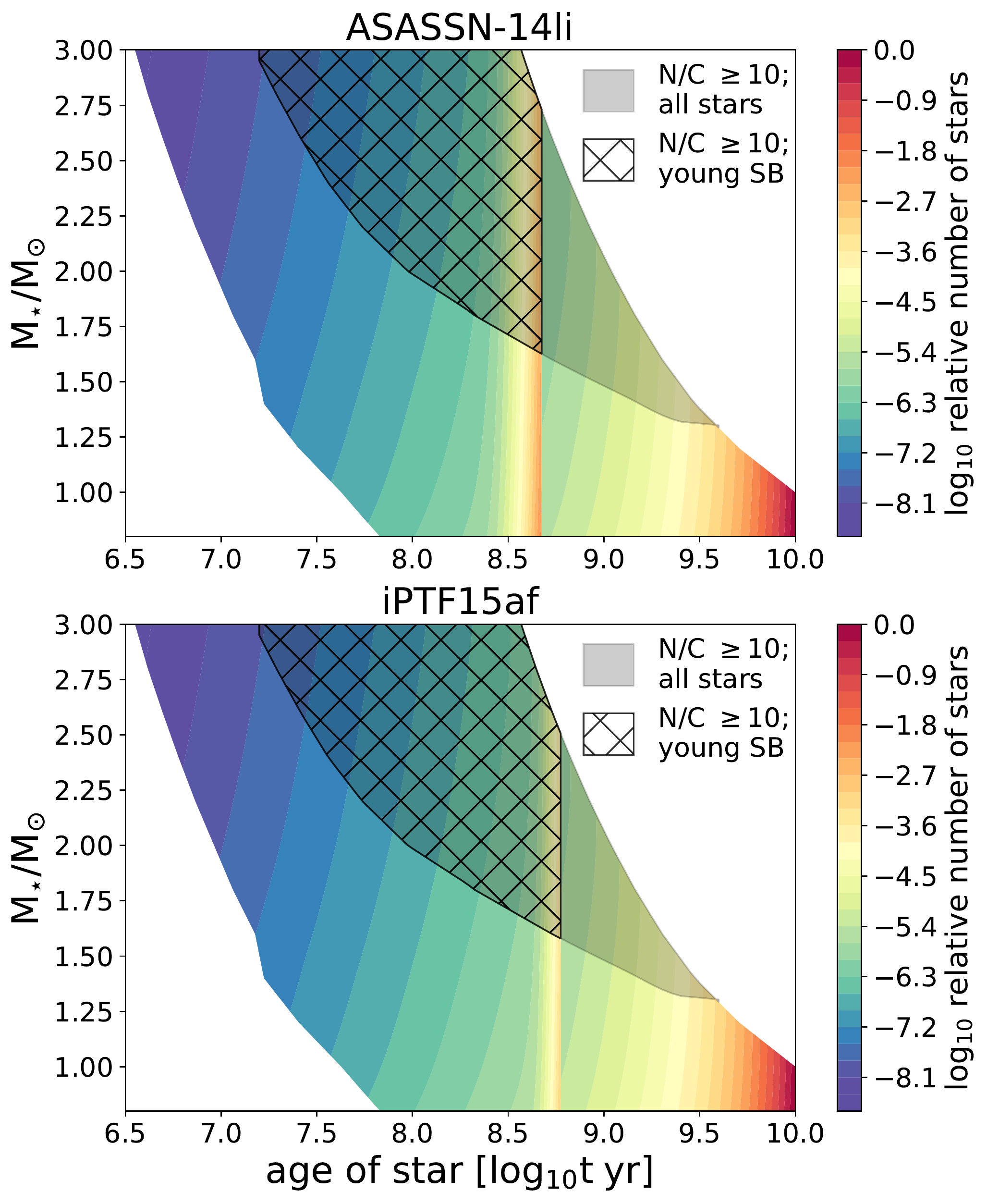} %\textwidth
\vspace{-0.3cm}
\caption{The relative number of main sequence stars in the host galaxies of ASASSN-14li and iPTF15af are plotted. We use the star formation rates for these galaxies as calculated by \citet{French:2017} to determine the relative population of young and old stars. We also weight the stellar population using the expected mass-dependent disruption cross section. We then calculate the N/C ratio in the tidal debris of each of these stars, using the formalism developed by \citet{Gallegos-Garcia:2018}. In the host of ASASSN-14li, the percentage of stars with N/C$\geq10$ is $0.028^{+0.008}_{-0.008}$\% in the entire stellar population of the host, and $8.2^{+1.4}_{-2.9}$\% in the recent starburst. In the host of iPTF15af, the percentage of stars with N/C$\geq10$ is $0.0045^{+0.0009}_{-0.0009}$\% in the entire stellar population, and $7.1^{+1.8}_{-0.7}$\% in the recent starburst. 
\label{fig:stellarpop}}
\vspace{-0.2cm}
\end{figure}

In Figure~\ref{fig:stellarpop} we plot the relative number of stars as a function of mass and age, and include contours defining the region where the N/C abundances are $\geq 10$. Most of the stars were formed $\approx10$ Gyrs ago while the recent starbursts in the hosts can be seen around 300 - 500 Myrs ago, where there is a significant increase in the number density of stars. In the host of ASASSN-14li, the percentage of stars in the entire stellar population that can produce tidal debris with N/C$\geq 10$ is $0.028^{+0.008}_{-0.008}$\% (the gray shaded region), although it rises to $8.2^{+1.4}_{-2.9}$\% if we only consider stars in the young starburst (the hatched region). In the host of iPTF15af, the percentage of stars is $0.0045^{+0.0009}_{-0.0009}$\% in the entire stellar population, and $7.1^{+1.8}_{-0.7}$\% when only considering stars within the recent starburst. 

In brief, the percentage of stars in the entire galactic population with sufficiently high N/C abundances to explain the observations is, as expected, very small.  However, it increases significantly if the  population of disrupted stars resembles those produced solely by the recent starburst. This would help to explain the observations without requiring 
a model that preferentially disrupts moderately massive stars (as we discuss further in Section~\ref{sec:disc}). 

It is also possible that the Kroupa IMF does not provide an accurate description for the distribution of stars in galactic nuclei. \citet{Lu:2013} measured the properties of the stellar population in the galactic center and found it to be best described  by a top-heavy IMF. They find a best fit power law `$\alpha$' index of $1.7^{+0.2}_{-0.2}$, compared to the Kroupa value of $\alpha = 2.3$ \citep{Kroupa:2001}. Substituting $\alpha = 2.3$ for $\alpha = 1.7$ for stars $\geq 0.5 M_\odot$, we find that the percentage of stars with N/C $\geq 10$ increases by a factor of $\approx 2$.
This brings the percentage to $0.06^{+0.02}_{-0.02} \%$ for the host galaxy of ASASSN-14li, and
to $0.009^{+0.002}_{-0.002} \%$ for the host galaxy of iPTF15af. Again, if we only look at the stars belonging to the recent starburst in each galaxy, the percentages are $15.2^{+2.9}_{-5.4} \%$ for the host of ASASSN-14li and $13.2^{+3.3}_{-1.3} \%$ for the host of iPTF15af.

Although there are no measurements of the star formation history for the host galaxy of iPTF16fnl, we can use the average age measured for the stellar population of the host to estimate the fraction of stars that fit our criteria. Two papers have measured the age of the host. \citet{Blagorodnova:2017a} estimates an age of $650^{+300}_{-300}$ Myr while \citet{Brown:2017a} estimates an age of $1.29^{+33}_{-0.3}$ Gyr. We estimate the percentage of stars with N/C $\geq 10$ for both ages, and find a range of 2.7-8.6\% using the Kroupa IMF, and 4.6-15.3\% using the \citet{Lu:2013} galactic center IMF. This of course assumes all stars in the galaxy formed at this age and as such it provides a strict upper limit.

\section{Discussion}\label{sec:disc}
\subsection{Assessing the ubiquity of N-rich TDEs}\label{sec:super-solar}

To determine the rarity of these moderately massive stellar disruptions, defined here as disruptions with $M_*\geq 1.3 M_\odot$, we must look at the entire TDE population. We have observations of $\approx $50 confirmed TDEs \citep{Auchettl:2018, van-Velzen:2021}. Only six TDEs have HST UV spectra, and of those, the three TDEs discussed in this paper ($\approx $6\% of the population) have high N/C abundance ratios as inferred from their UV spectra.  There are $\sim 10$ other TDEs that have nitrogen lines present in their optical spectra but lack UV spectra (such as ASASSN-18pg). It is certainly possible that these events are also nitrogen-rich, which makes the derived fraction of 6\% a strict lower limit. Nevertheless, 6\% is well in excess of the fraction of stars that could produce these nitrogen abundances ($\approx 0.0005\%-0.06\%$) if the populations of disrupted stars is selected randomly from the total stellar population of the host. The fraction of stars able to produce these nitrogen abundances drastically increases if the disrupted stars are drawn from the  recent starburst population. Motivated by this, we look more closely at the percentage of moderately massive disruptions among TDEs in galaxies with recent starbursts, instead of simply looking at the percentage of of these disruptions among all TDEs.

In a recent review paper, \citet{French:2020a} found that of the 41 TDE candidates with derived host galaxy properties, 5 of them were discovered in post-starburst galaxies and 13 were discovered in either quiescent Balmer-strong or post-starburst galaxies.  This implies that 2/5 (40\%) of TDEs observed in post-starburst galaxies and 3/13 (23\%) of TDEs observed in quiescent Balmer-strong galaxies have enhanced nitrogen abundances, as inferred  by their UV spectra. This is still higher than the fraction of stars that could produce the measured nitrogen abundances in these galaxies, even if all of the disrupted stars came from the recently formed population of stars ($\approx 7 - 15\%$ for iPTF15af and ASASSN-14li). What is more, this is a lower limit, given that there is not UV spectra for the other events in QBS galaxies. As such, this implies an enhanced disruption rate of moderately massive stars (key to determining the mechanism responsible for transporting stars near the SMBH). 

\subsection{Theories for TDE enhancement}

Several theories have been put forward to explain the enhanced TDE rate in post-starburst galaxies, and it is possible that these might also help explain the overabundance of moderately massive stellar disruptions in post-starburst galaxies. The TDE enhancement can be ascribed to a dynamical mechanism that preferentially transports and disrupts moderately massive stars and/or to an overabundance of moderately massive stars in galactic nuclei.

\subsubsection{Star formation in galactic nuclei}
One possible explanation is that the general stellar population of stars in a starburst galaxy is not representative of that in its nucleus. For example, \citet{Norton:2001, Yang:2008a} found that in E+A galaxies the Balmer absorption features become more prominent as one moves radially inward. Building on this work, \citet{Pracy:2012, Pracy:2013} found that the young stellar populations in nearby E+A galaxies have centrally concentrated gradients. This matches what simulations predict for a starburst caused by a gas-rich merger, as cold gas is funnelled into the center of the galaxy, triggering a nuclear starburst \citep{Mihos:1994, Bekki:2005, Hopkins:2009a}.  Interestingly, we also observe elliptical galaxies with blue cores \citep{Menanteau:2001, Pipino:2009}, and dwarf galaxies that possibly experienced nuclear starbursts \citet{Sanchez-Janssen:2019}, which indicates that the disruption of moderately massive stars can still take place in galaxies with no apparent star formation. The reader is referred to the discussion of the TDE ASASSN-18pg in Appendix~\ref{app:fwhm}.

As previously discussed in Section~\ref{sec:hostgalaxies}, an increased fraction of young stars in the cores of these galaxies can help explain the overabundance of TDEs arising from the disruption of moderately massive stars. However, in some cases, there are not enough super-solar stars to explain observations even when assuming that all of the stars in the core are from the galaxy's most recent starburst. In Section~\ref{sec:hostgalaxies} we also showed that a further enhancement can be produced if the IMF of stars in galactic nuclei is top-heavy (compared to a Kroupa IMF), as inferred for the Milky Way's galactic center \citep{Lu:2013} and predicted in extreme star forming environments \citep{Larson:2005, Hennebelle:2008,Bate:2009a}. Unfortunately, it is difficult to study this observationally -- the lowest mass stars we are currently able to observe in our own galactic center are $\gtrsim 2 M_\odot$ \citep{Hosek:2019b}. If we assume, for example, a low-mass cutoff of $\gtrsim 0.5 M_\odot$ in addition to using a top-heavy IMF slope and assuming that stars in galactic nuclei are drawn from the young stellar galactic population, then the percentage of stars with N/C abundances $\gtrsim10$ that are available to be disrupted in the hosts of iPTF15af and ASASSN-14li would increase to around $\approx 20$\%. This would be in close agreement with observational constraints. JWST will be key for testing these theories as it should be able to resolve $\approx 0.3 M_\odot$ stars in the Arches cluster near the galactic center \citep{Hosek:2019b}. 

It is also possible that stars in galactic nuclei are younger than those associated with the overall starburst episode. The Arches cluster in the galactic center, for example, underwent recent star formation \citep[around ~2-5 Myr ago,][]{Paumard:2006a, Negueruela:2010, Gennaro:2011, Lu:2013}, and measurements of the star formation history in the galactic center are consistent with quasi-continuous star formation \citep{Schodel:2020}. The conditions in the galactic nuclei are thought to be similar to the conditions in highly star forming starburst episodes \citep{Ginsburg:2016a}. This would obviously increase the number of moderately massive stars available for disruption, although it would not necessarily increase the percentage of stars with high N/C abundances, as stars below $\approx  2 M_\odot$ take $\approx  100$ Myr to develop such enhancements (see Figure~\ref{fig:stellarpop}). 

\subsubsection{Dynamical mechanisms leading to TDEs}
Another potential explanation for the observed TDE enhancement is that the leading dynamical mechanisms at work in these galactic nuclei are preferentially increasing the rate of high mass disruptions. Mass segregation has been observed in star clusters for stars above $\approx 5 M_\odot$ \citep{Hillenbrand:1998, Gennaro:2011}, although it is unclear whether this can be solely explained by two-body relaxation or if the actual star formation process needs to be altered \citep{Parker:2011, Dib:2018,Plunkett:2018}. Two-body relaxation is commonly assumed to be responsible for moving stars onto loss-cone  orbits \citep{Wang:2004a, Stone:2016a}, and for galaxies where it is the dominant mechanism for producing TDEs, some mass segregation is to be expected \citep{Allison:2009b}. Calculating the dynamics of a non-homogeneous population of stars requires N-body simulations, and there is not a simple analytical adjustment to the rates as a function of stellar mass. Simulations have found that the highest mass stars generally follow a Bahcall-Wolfe density profile ($r^{-\alpha}$; $\alpha \approx 1.75$, \citealt{Bahcall:1977a}), while the total stellar mass follows a shallower profile ($\alpha \approx 1.5$, \citealt{Baumgardt:2004, Vasiliev:2017}). Direct N-body simulations of multi-mass star clusters containing intermediate-mass black holes by \citet{Baumgardt:2004} found that $\alpha$ increased from 0.7 for $\approx 0.1 M_\odot$ stars to 1.7 for $\approx  1 M_\odot$ stars, so that the average mass of stars in the core was $\approx  0.6 M_\odot$ and their simulations show a clear preference for the disruption of $\gtrsim 0.6 M_\odot$ stars. Yet, it remains unclear if mass segregation in galactic nuclei could explain the observed preference for moderately massive stellar disruptions.  

Another dynamical alternative is that these galactic nuclei host eccentric nuclear disks \citep[such as the one in Andromeda,][]{Tremaine:1995}. \citet{Madigan:2018} and \citet{Wernke:2019} found that eccentric nuclear disks, which could originate in galaxy mergers \citep{Hopkins:2010a}, can dramatically increase the rate of tidal disruption events. Galaxies that experienced recent starbursts are good candidates for recent galaxy mergers, as mergers funnel gas to the centers of galaxies and drive up star formation \citep{Barnes:1991a}. Additionally, \citet{Foote:2020} found that radial mass segregation takes place within eccentric nuclear disks,  implying that they would naturally produce a higher fraction of high mass stellar disruptions than would be predicted from the underlying stellar population. Therefore, eccentric nuclear disks would not only increase the total rate of disruptions in post-merger galaxies, but could also help explain the relative rate of disruptions of moderately massive stars. 

Two other dynamical mechanisms that could help increase the rate of tidal disruptions in post-merger galaxies are the Kozai-Lidov (KL) mechanism and the eccentric Kozai-Lidov (EKL) mechanism \citep{Chen:2011a,Li:2015a}. These mechanisms are a natural consequence of the presence of a SMBH binary assembled after a galaxy merger. A star orbiting one of the SMBHs will be perturbed by the second SMBH, and this perturbation can cause the stellar eccentricity and inclination to undergo periodic oscillations, potentially ending in the star being disrupted. Both the KL and EKL mechanisms have been explored as ways to increase the total number of disruptions in post-starburst galaxies \citep{Stone:2018a, Mockler:2021c}. Both mechanisms can help increase the rate of TDEs after a merger, but their efficacy depends on stars within the sphere of influence of the SMBH binary being effectively replenished through, e.g., regulated star formation. Neither of these mechanisms preferentially disrupt higher mass stars on their own, however if they are combined with top-heavy star formation or the formation of nuclear disks, it is possible they could explain both the enhanced rate of TDEs in post-starburst galaxies and the overabundance of moderately massive disruptions in these galaxies. 

Regardless of the true mechanism (or mechanisms) for increasing the rate of moderately massive tidal disruptions, we expect our understanding of nuclear stellar clusters will continue to be challenged and enhanced by TDE discoveries.

\begin{acknowledgments}
We thank Tiara Hung, Giorgos Leloudas, Nadia Blagorodnova and John Brown for their helpful collaboration on this work. We thank the anonymous referee for constructive corrections and suggestions. B.M. is grateful for the AAUW American Fellowship, and the UCSC Presidents Dissertation Fellowship. A.A.T., E.R.-R. and B.M. are grateful for support from the Packard Foundation, Heising-Simons Foundation, NSF (AST-1615881, AST-1911206 and AST-1852393), Swift (80NSSC21K1409, 80NSSC19K1391) and Chandra (GO9-20122X). 
K.A. is grateful for support from the Australian Research Council Centre of Excellence for All Sky Astrophysics in 3 Dimensions (ASTRO 3D), through project number CE170100013. 
The calculations for this research were carried out in part on the UCSC Hyades supercomputer (supported by NSF award AST-1229745 and by UCSC), and in part on the UCSC lux supercomputer (supported by NSF MRI grant AST-1828315). 

\end{acknowledgments}
\vspace{5mm}

\software{\mosfit \citep{Guillochon:2017a}, astropy \citep{Astropy-Collaboration:2013a}
}

\bibliography{library}{}
\bibliographystyle{aasjournal}

\appendix
\section{TDE ASASSN-18pg}
\label{app:18pg}
\begin{figure}[ht!]
\centering 
\begin{center}
\includegraphics[scale = 0.52]{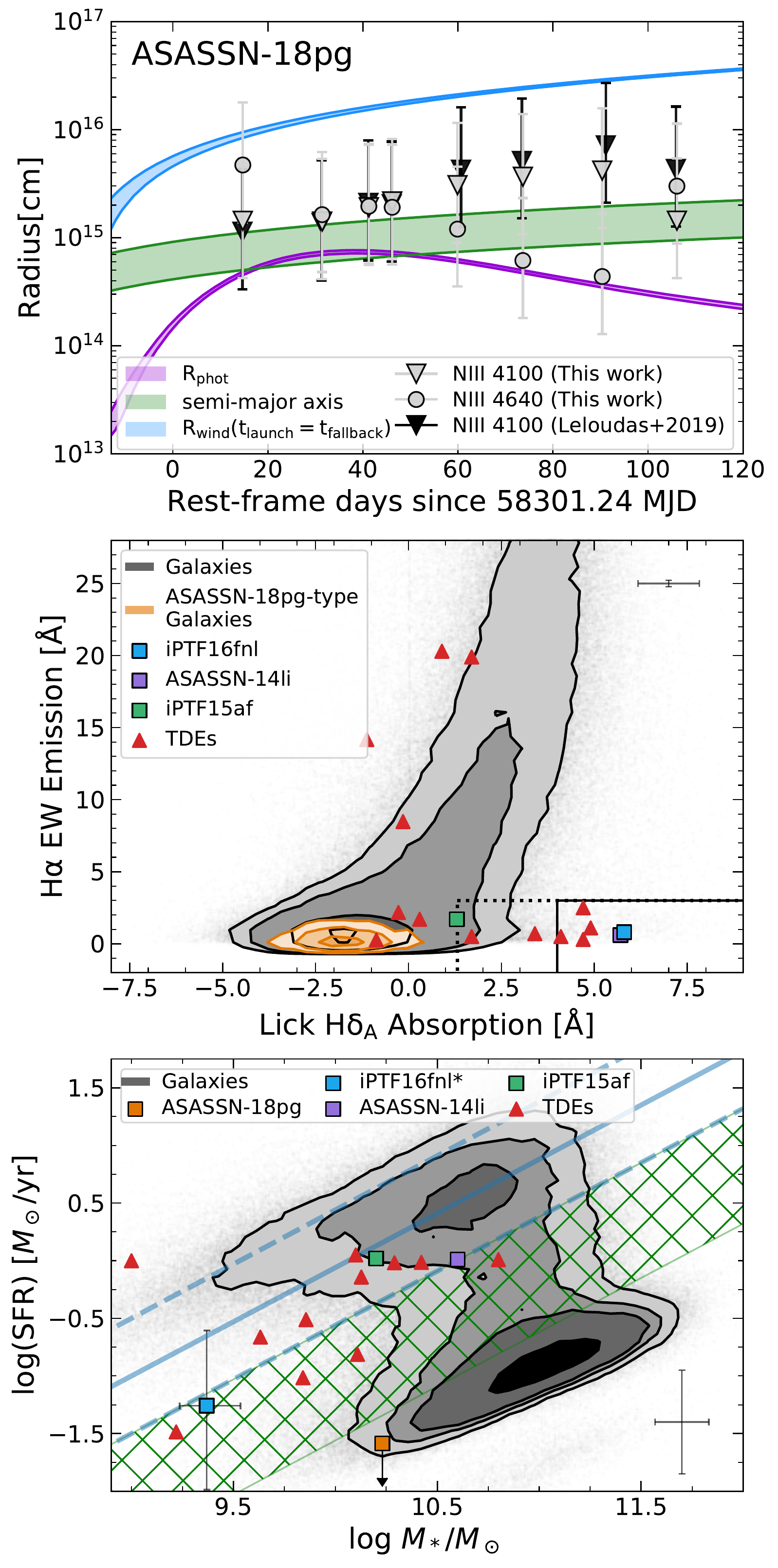}
\end{center}
\vspace{-0.5cm}
\caption{Analogous to Figures~\ref{fig:blr} and~\ref{fig:hostgal}, see relevant figure captions for further description. {\bf Top panel:} The blackbody photosphere is plotted in purple, the semi-major axis of debris returning at a given time is plotted in green, and the wind radius is plotted in blue. {\bf Middle panel:} H$\alpha$ equivalent width as a function of Lick H$\delta$A absorption. Galaxies inside the solid line box are E+A/PSB, galaxies inside the dashed line box are QBS. Because there are not measurements of H$\alpha$ EW or Lick H$\delta$A for the host of ASASSN-18pg, we estimated their likely range by taking the distribution of these parameters for galaxies with similar values of $M_\ast$ and SFRs. We used galaxies within 1$\sigma$ in stellar mass that had star formation rates consistent with the upper limit for the host of ASASSN-18pg. {\bf Bottom panel:} The star formation rate as a function of the total stellar mass.
\label{fig:18pg}}
\end{figure}

The TDE ASASSN-18pg has optical nitrogen lines observable through much of its evolution \citep{Leloudas:2019}. We find that these nitrogen lines are consistent with originating from the tidally disrupted star (Figure~\ref{fig:18pg}, top panel), although without UV spectra it is difficult to determine how the line strength is connected to the star's composition. Unlike the other TDEs analyzed in this paper, ASASSN-18pg was not discovered in a post starburst galaxy but in an elliptical one  (Figure~\ref{fig:18pg}, bottom panels). We note that there are $\approx 10$ TDE candidates (including ASASSN-18pg) that have nitrogen measured in their optical spectra but lack UV spectra. Four of these have host galaxy classifications, and of these, one occurred in a post starburst galaxy and three were found in quiescent galaxies. If the associated nitrogen lines are due to the disruption of a moderately massive star, then this implies that the cores of these quiescent galaxies contain younger stars. This is consistent with the seminal results from \citet{Menanteau:2001} showing that some quiescent galaxies have blue cores. 

\section{nitrogen emission line analysis}\label{app:fwhm}
To constrain the size scales of the nitrogen rich material, we take advantage of the publicly available optical spectra for ASASSN-14li \citep{Holoien:2016a, Cenko:2016, Brown:2017a}, iPTF15af \citep{Blagorodnova:2019, Onori:2019}, iPTF16fnl \citep{Blagorodnova:2017a, Brown:2018} and ASASSN-18pg \citep{Leloudas:2019,Holoien:2020}. Here we constrain the radii and evolution of this gas using the full-width half max (FWHM) of the nitrogen emission lines detected in these events.  As discussed in e.g., \citet{Holoien:2020, Charalampopoulos:2021}, line profile fitting can be complicated due to the broadness of the lines, blending of nearby emission lines, the signal to noise of the spectra, and the different methods of analysis presented in the literature. As such, rather than taking solely taking the FWHM of these lines from the literature, we re-analyse the spectra of each source using a consistent method as presented below.  

As we are most interested in the emission lines of these events, we first model the continuum emission of each spectrum using the Astropy package, \textit{specutils\footnote{\url{https://specutils.readthedocs.io/en/stable/}}}. The spectra is smoothed using a median filter to minimise any extreme variations in the spectra. The continuum is then fitted using a low-order Chebyshev polynomial of the 1st kind. To obtain the normalised spectra which we use for our analysis, we divided our smoothed spectra by our best fit continuum model. This method allows us to better determine the evolution of these lines in terms of their equivalent width and strengths relative to continuum. As we do not correct our spectra for host contribution, we note that this method may not fully remove host emission and this is one of the main sources of our systematic errors. Note that \citet[e.g.,][]{Holoien:2020} suggest that the uncertainty in identifying and removing continuum emission can correspond to over 25\% of the error estimate.

To constrain the FWHM of the nitrogen lines (and other lines present in the spectrum between 3900\AA and 5200\AA), we fit each emission line profile using a simple Gaussian, similar to what is done in the literature \citep[e.g.,][]{Hung:2017a, Blagorodnova:2019, Leloudas:2019}. For \ion{N}{3} (4640 \AA), which is blended with He II (4686 \AA), we fit both lines simultaneously. While additional components or the use of a Lorentizian model may be more physically motivated or better describe the data \citep[see ][for more discussion]{Charalampopoulos:2021}, a simple Gaussian model is sufficient for our purposes. 

\begin{figure}[ht!]
\centering 
\begin{center}
\includegraphics[width=0.45\textwidth]{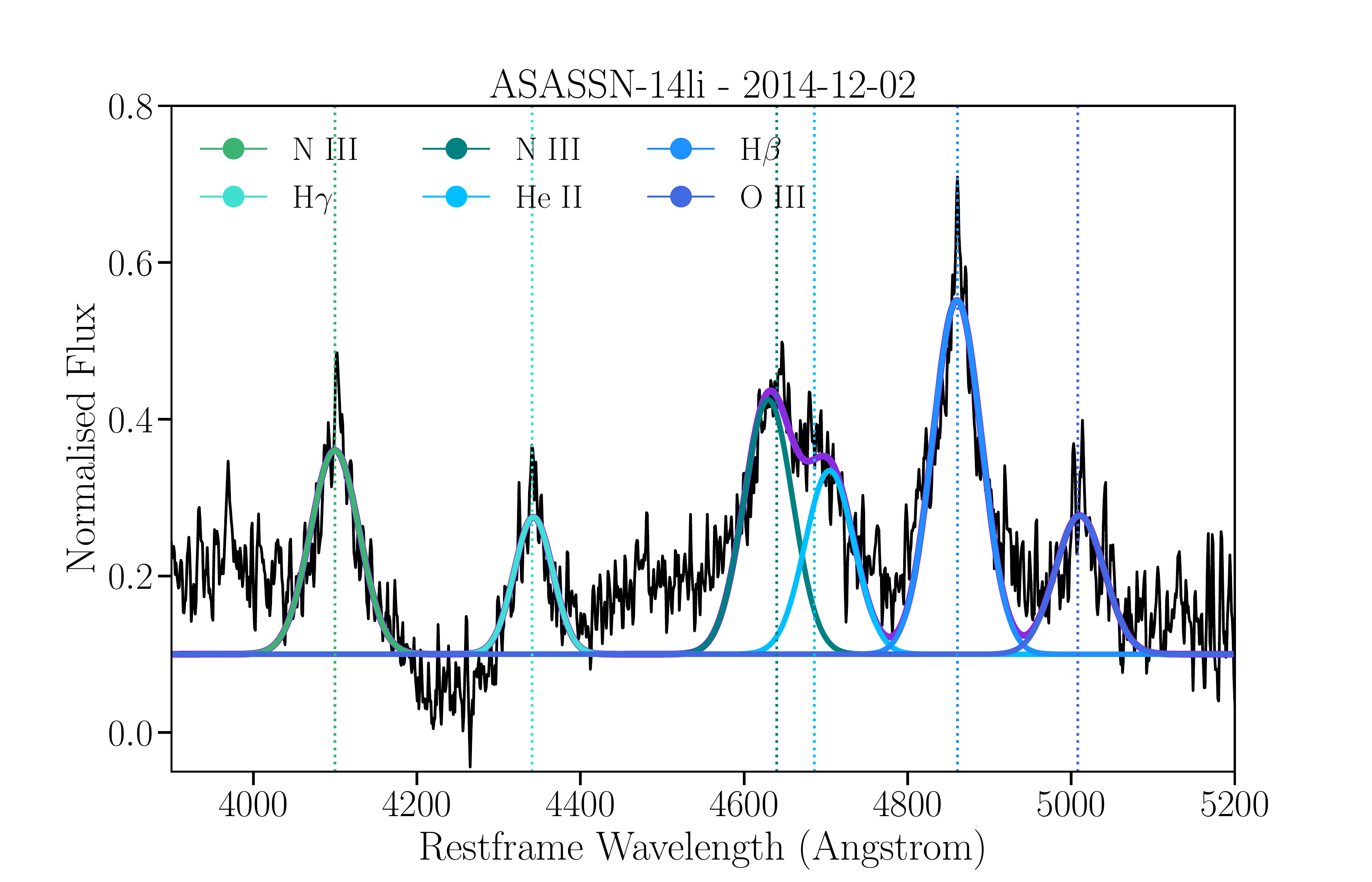}
\includegraphics[width=0.45\textwidth]{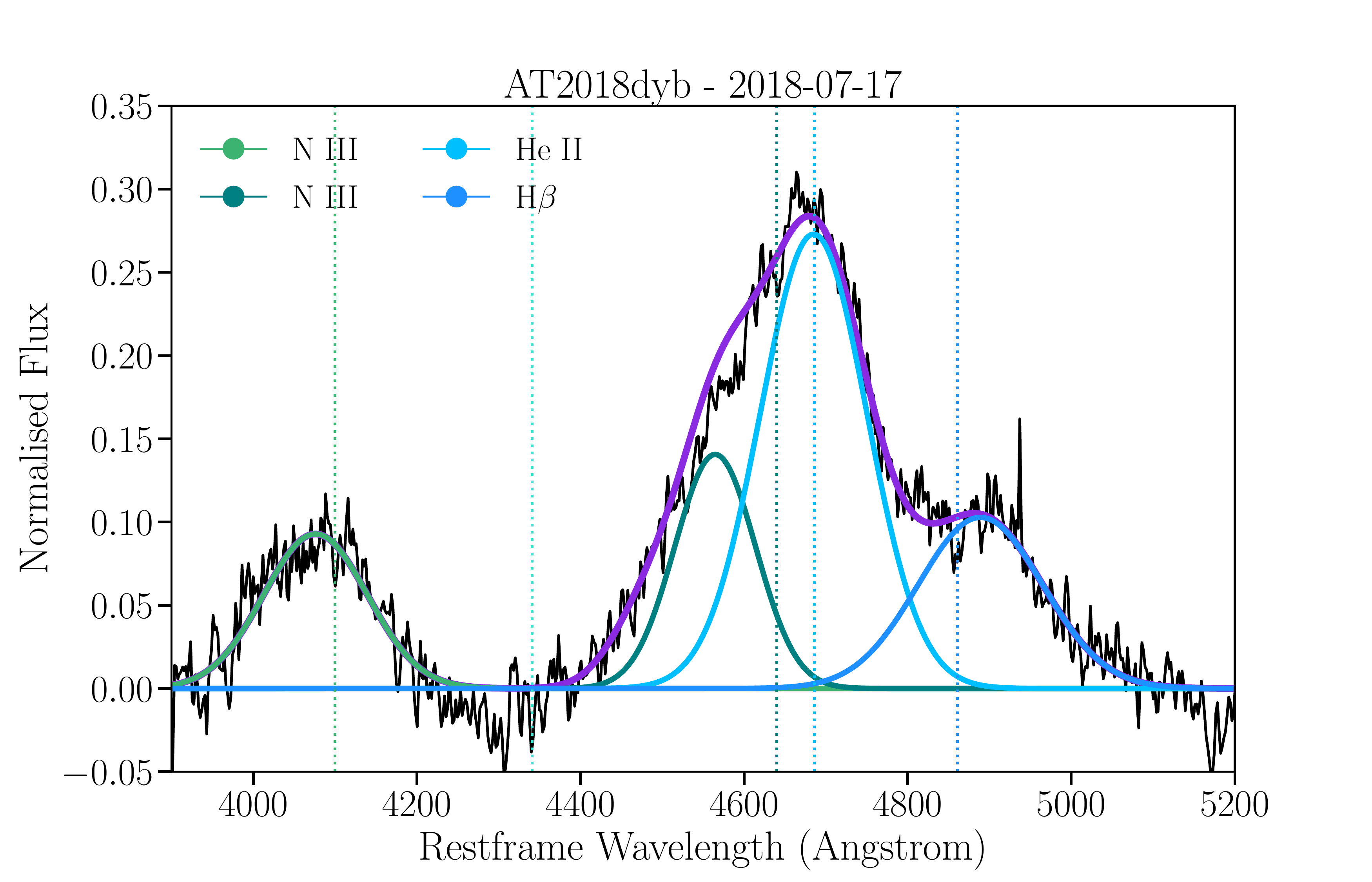}
\includegraphics[width=0.45\textwidth]{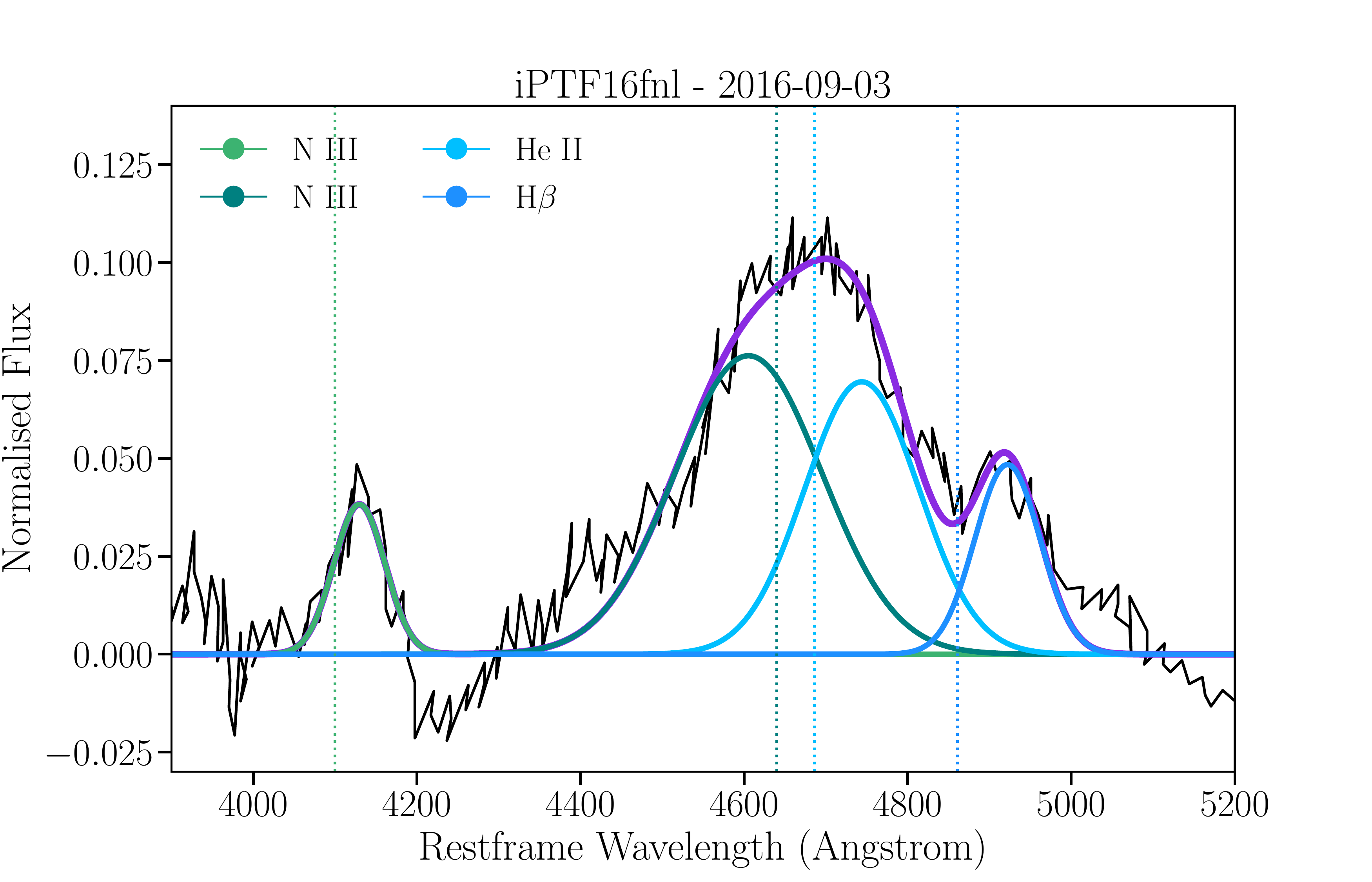}
\includegraphics[width=0.45\textwidth]{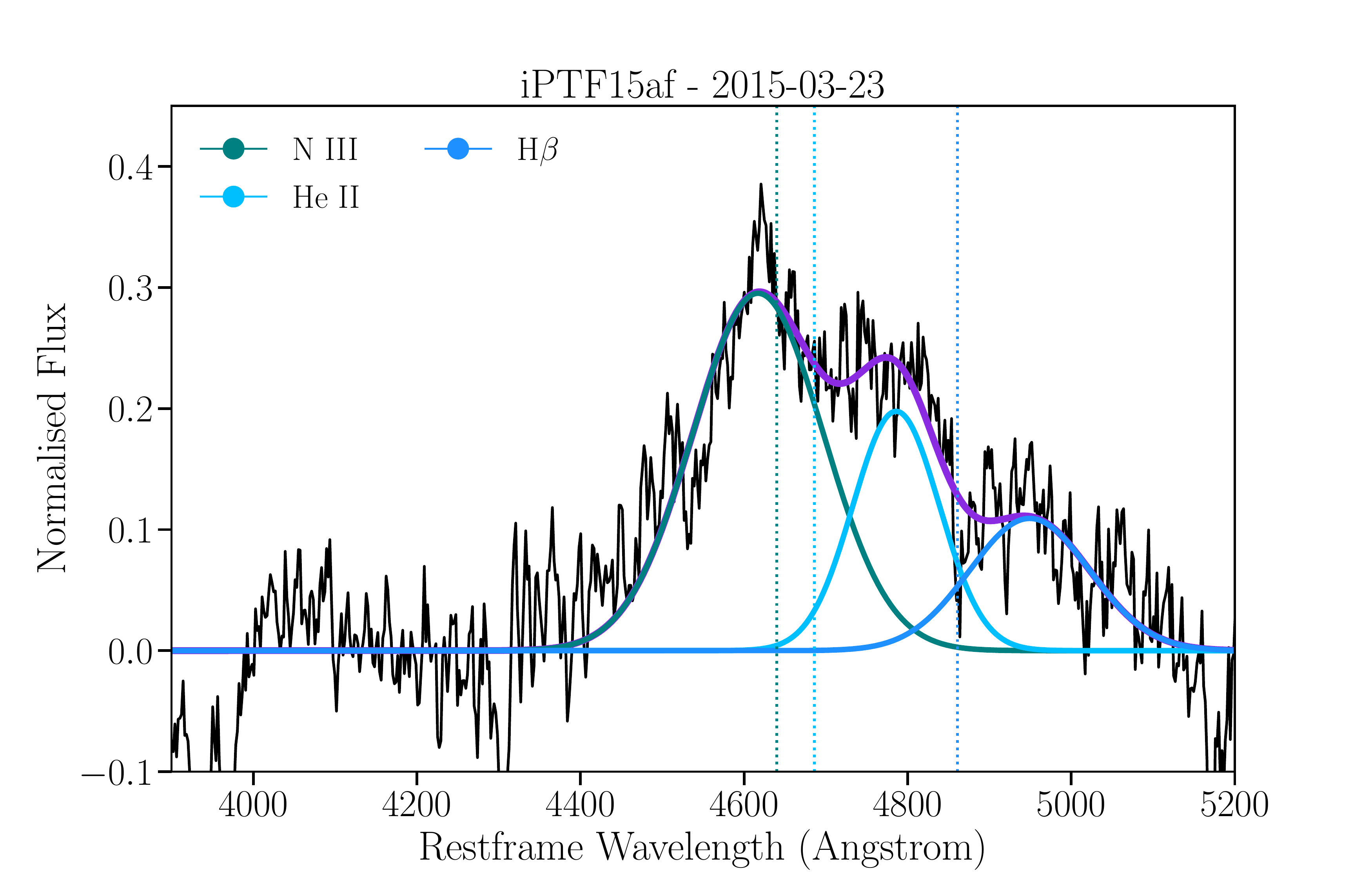}
\end{center}

\caption{Example best fit models obtained by simultaneously fitting multiple Gaussians in the spectrums of ASASSN-14li (taken using APO 3.5m on 2014-12-02), AT2018dyb (taken using SOAR on 2018-07-17), iPTF16fnl (taken using the Nordic Optical Telescope on 2016-09-03) and iPTF15af (taken using Keck LRIS on 2015-05-23), respectively. The central wavelengths of each emission lines (assuming zero velocity) are marked by vertical dotted lines. 
\label{fig:fhwm}} 

\end{figure}

To perform this fit, we use Markov Chain Monte Carlo (MCMC) methods. Here we let the amplitude of the emission profile, standard deviation, and the wavelength of the center of the peak vary, but set uniform priors for each parameter to ensure convergence. Using the best fit values, we derive the velocity width or FWHM (and the velocity shift from the rest wavelength) for each line and derive 90\% confidence error intervals for these parameters. In Figure \ref{fig:fhwm} we show examples of the best fit models derived from our analysis of each event, while in Table \ref{tab:fhwm} we list the best fit FWHM for each nitrogen line and their uncertainties. We find that our method produces similar results to those found in the literature (see Figure \ref{fig:blr} and \ref{fig:18pg}).

\begin{table}[]
    \centering
        \caption{The derived emission line widths of N III 4100\AA\, and N III 4640\AA\, for each tidal disruption event in our sample.}
    \label{tab:fhwm}
\resizebox{\textwidth}{!}{    \begin{tabular}{llcccccccccc}
    \hline
	&		&		&		&	N IIII  4100	&	N IIII  4100	&	N III 4640	&	N III 4640	\\
TDE	&	Spectrum	&	Data of observation	&	Days since TDE discovery	&	FWHM (km/s)	&	FWHM uncertainty	&	FWHM (km/s)	&	FWHM uncertainty	\\
\hline \hline
ASASSN-14li	&	APO 3.5m	&	2014-12-02	&	10	&	3300	&	81	&	6317	&	4069	\\
ASASSN-14li	&	MDM 2.4m	&	2014-12-09	&	17	&	$-$	&	$-$	&	6019	&	406	\\
ASASSN-14li	&	MDM 2.4m	&	2014-12-10	&	18	&	$-$	&	$-$	&	5740	&	1805	\\
ASASSN-14li	&	APO 3.5m	&	2014-12-12	&	20	&	3691	&	66	&	6085	&	237	\\
ASASSN-14li	&	FLWO 1.5m	&	2014-12-14	&	22	&	3939	&	19	&	6074	&	1054	\\
ASASSN-14li	&	FLWO 1.5m	&	2014-12-15	&	23	&	3444	&	2192	&	6047	&	835	\\
ASASSN-14li	&	FLWO 1.5m	&	2014-12-19	&	27	&	4776	&	16	&	6070	&	491	\\
ASASSN-14li	&	MDM MODS 2.4m	&	2014-12-23	&	31	&	1984	&	23	&	2812	&	42	\\
ASASSN-14li	&	Magellan Baade	&	2015-01-03	&	42	&	1652	&	19	&	1865	&	46	\\
ASASSN-14li	&	LBT MODS	&	2015-01-20	&	59	&	2113	&	277	&	1508	&	26	\\
ASASSN-14li	&	MDM 2.4m	&	2015-02-11	&	81	&	$-$	&	$-$	&	7572	&	463	\\
ASASSN-14li	&	LBT MODS	&	2015-02-16	&	86	&	2099	&	251	&	3367	&	881	\\
ASASSN-14li	&	MDM 2.4m	&	2015-03-11	&	109	&	$-$	&	$-$	&	$-$	&	$-$	\\
ASASSN-14li	&	LBT MODS	&	2015-04-17	&	146	&	$-$	&	$-$	&	2209	&	27	\\
\hline
iPTF15af	&	Keck LRIS	&	2015-01-22	&	7	&	$-$	&	$-$	&	13009	&	267	\\
iPTF15af	&	FLOYDS North	&	2015-02-01	&	17	&	$-$	&	$-$	&	22728	&	2024	\\
iPTF15af	&	Keck LRIS	&	2015-02-23	&	67	&	$-$	&	$-$	&	12240	&	176	\\
iPTF15af	&	Keck LRIS	&	2015-05-16	&	121	&	$-$	&	$-$	&	12448	&	73	\\
iPTF15af	&	Keck LRIS	&	2015-06-13	&	149	&	$-$	&	$-$	&	12422	&	862	\\
iPTF15af	&	Keck LRIS	&	2017-11-15	&	1035	&	$-$	&	$-$	&	12447	&	392	\\
\hline
iPTF16fnl	&	Nordic Optical Telescope	&	2016-08-31	&	5	&	10238	&	2750	&	9960	&	1133	\\
iPTF16fnl	&	Nordic Optical Telescope	&	2016-09-03	&	8	&	5377	&	3617	&	13802	&	473	\\
iPTF16fnl	&	Nordic Optical Telescope	&	2016-09-09	&	14	&	10332	&	103	&	10603	&	373	\\
iPTF16fnl	&	Nordic Optical Telescope	&	2016-09-13	&	18	&	10404	&	43	&	10618	&	296	\\
iPTF16fnl	&	Nordic Optical Telescope	&	2016-09-21	&	26	&	10359	&	366	&	10815	&	1455	\\
iPTF16fnl	&	Nordic Optical Telescope	&	2016-10-03	&	38	&	10338	&	76	&	10819	&	255	\\
iPTF16fnl	&	Nordic Optical Telescope	&	2016-10-11	&	46	&	10440	&	6686	&	10453	&	4485	\\
iPTF16fnl	&	Nordic Optical Telescope	&	2016-10-31	&	66	&	10406	&	7058	&	10621	&	5226	\\
\hline
AT2018dyb	&	SOAR Goodman	&	2018-07-17	&	6	&	10994	&	294	&	6066	&	2632	\\
AT2018dyb	&	Nordic Optical Telescope 	&	2018-08-03	&	23	&	11076	&	1062	&	10295	&	5057	\\
AT2018dyb	&	Nordic Optical Telescope 	&	2018-08-13	&	33	&	9519	&	404	&	9413	&	820	\\
AT2018dyb	&	Nordic Optical Telescope 	&	2018-08-18	&	38	&	8949	&	373	&	9518	&	1384	\\
AT2018dyb	&	Nordic Optical Telescope 	&	2018-09-01	&	52	&	7520	&	433	&	12008	&	4046	\\
AT2018dyb	&	Nordic Optical Telescope 	&	2018-09-15	&	66	&	6875	&	273	&	16784	&	7261	\\
AT2018dyb	&	Nordic Optical Telescope 	&	2018-10-02	&	83	&	6439	&	2192	&	19905	&	10359	\\
AT2018dyb	&	Nordic Optical Telescope 	&	2018-10-18	&	99	&	10981	&	6954	&	7597	&	4030	\\
\hline \hline
    \end{tabular}
    }

\end{table}

\end{document}